\documentstyle[12pt,epsfig]{article}

\def\beq{\begin{equation}}
\def\eeq{\end{equation}}
\def\ber{\begin{eqnarray}}
\def\eer{\end{eqnarray}}

\def\a+{\alpha_+}
\def\beqn{\begin{eqnarray}}
\def\eeqn{\end{eqnarray}}

\begin{document}

\baselineskip=15.5pt \pagestyle{plain} \setcounter{page}{1} 
\begin{titlepage}


\begin{center}

\vskip 6.7 cm

{\LARGE {\bf {Correlators in AdS$_3$ string theory}}}

\vskip 2.1 cm

{\large Gast\'on Giribet and Carmen N\'u\~nez}

\vskip 1.0 cm

Instituto de Astronom\'{\i}a y F\'{\i}sica del Espacio

C.C. 67 - Suc. 28, 1428 Buenos Aires, Argentina

and

Physics Department, University of Buenos Aires

gaston, carmen@iafe.uba.ar

\vskip 2.1 cm

{\bf Abstract}
 \end{center}

The computation of two and three point functions in the Coulomb gas free
field approach to string theory in the SL(2,R)/U(1) black hole background 
is reviewed. An interesting relation arises when
comparing the results obtained using two different
screening operators.
The formalism is then modified to study string theory propagating in
AdS$_3$ which
is considered as the direct product of
the SL(2)/U(1) coset times a timelike free boson. This representation
allows to naturally include the spectral flow symmetry and winding
number in vertex operators and correlation functions. Two and three
point tachyon amplitudes are computed in this new scenario and the results
coincide with
previous reports in the literature. Novel expressions are found for
processes violating winding number conservation.

\end{titlepage}

\newpage

\section{Introduction}

String theory on  three dimensional Anti de Sitter space (AdS$_3$) is an
interesting model to analyse the AdS/CFT correspondence beyond the field
theory approximation. Much progress has been achieved in understanding this
theory in recent years (see \cite{todos,hwang}), although there are still
important
issues to be clarified.

Unitarity has been the leading subject in this story during the last
decade
since, unlike string theory in flat spacetime, the Virasoro constraints
seemed unable to annihilate all the negative norm states of the string
propagating in AdS$_3$. Fortunately a naturally unitary spectrum has been
revealed by the spectral flow symmetry disclosed in reference \cite{mo}.
The
new representations obtained by the spectral flow were originally considered
in \cite{hwang} in the context of string theory in the SL(2,R) group manifold.
It was shown in reference \cite{mo} that they resolve some of the 
longstanding 
negative consequences of  arbitrarily truncating the spin $j$
(equivalently the mass) of the physical states when the traget
space is the universal cover of SL(2,R). This represents an important
step in the construction of a consistent model, but the consistency of the
theory cannot be completely established
until interactions are included and the closure of the operator product
expansion is determined. Indeed, regarded as a conformal field theory,
string theory in AdS$_3$ will be completly characterized by the spectrum and
the full set of three point functions.

In this paper we continue the study of correlation functions of string
theory in AdS$_3$ which was started in reference \cite{gn2}. Based on the
proposal in \cite{mo} we consider the theory as the tensor product of the
coset space $H^+_3/U(1)$ (the euclidean version of the SL(2,R)/U(1) group
manifold) times the state space of a timelike free boson. The vertex
operators are constructed and correlation functions are computed
extending to the
non-compact case a prescription developed for SU(2) in reference 
\cite{dotse}.
 This formalism 
is suitable to manifestly include the spectral flow
parameter or winding number $\omega$. We explicitly construct two and three
point functions of physical states using the modified Coulomb gas
formalism developed in \cite{gn2},
and we then compare the expressions obtained with results reported earlier 
in the literature which were found by
other methods. 

Actually, various approaches have been followed to construct correlation
functions in
the \- SL(2,R)\- WZW \- model (the lagrangian formalism \cite{gawe,
satoh}, the
bootstrap method \cite{tesch1, teschner}, the free field approximation \cite
{morozov, becker, satoh2}). The results obtained in these references
were shown to agree in
 \cite{satoh2}, where it is argued furthermore that the full
interacting theory might be reducible to a free theory based on the
precise equivalence of two, three and certain four point functions.
Here we confirm that this is indeed the case for correlation functions of
two
and three tachyon states, using a different free field approach.

The plan of the article is as follows. In Section 2 we review the
computation of two and three point tachyon
amplitudes
in the SL(2,R)/U(1) two dimensional black hole background performed in
reference \cite{becker} using the Coulomb gas free field approximation. As
it is well known, there are two screening operators in this formalism
\cite
{ooguri, andreev}, so we take the opportunity to compare the results
obtained using each of them. This leads to an interesting relation,
similar
to one recently proposed in Liouville theory \cite{tesch3}. In Section 3
we recompute the two and three point tachyon amplitudes using a modified
Coulomb
gas
prescription, suitable to deal with string theory in AdS$_3$.
The results are shown to agree with previous reports in the
literature for processes conserving winding number. However, novel results
are
obtained for three point functions violating winding number conservation
by one
unit. A summary of the results accomplished and conclusions are contained
in Section 4.

\section{Free field approach to SL(2,R) WZW model}

This Section contains a review of the computation of correlation functions
in SL(2,R) CFT using the Coulomb gas prescription and a detailed analysis
of
the ingredients that are needed in the standard formalism. Since we are
ultimately interested in the application of the techniques to string
theory in AdS$_3$, we start by briefly recalling the free field
formulation
of this model.

 String theory on
$AdS_3$ is described by the lagrangian \begin{equation}
L = k(\partial\phi\bar\partial\phi + e^{2\phi}\bar
\partial\gamma\partial\bar\gamma)  \label{lagran}
\end{equation}
where $(\phi, \gamma, \bar\gamma)$ are coordinates on the Euclidean $AdS_3$
spacetime, which is equivalent to the quotient space $H_3^+ = SL(2,C)/SU(2)$%
, and $k$ is a constant related to the radius of curvature of spacetime $l$
and the fundamental string length $l_s$, as $k = l^2/l_s^2$. Conformal
invariance of the sigma model on this background requires in addition a
NS-NS antisymmetric tensor field.

It is convenient to rewrite the action adding one-form auxiliary fields
$\beta$, $\bar
\beta$ as 
\begin{equation}
L = k(\partial\phi\bar\partial\phi + \beta\bar\partial\gamma +
\bar\beta\partial\bar\gamma - e^{-2\phi}\beta\bar\beta),
\end{equation}
which gives (\ref{lagran}) after integrating out $\beta$, $\bar\beta$.

This theory is a SL(2) WZW model with a current algebra specified by the
following
OPE 
\begin{eqnarray}
J^{+}(z)J^{-}(w) &=&\frac k{(z-w)^2}-\frac 2{(z-w)}J^3(w)+...  \nonumber \\
J^3(z)J^{\pm}(w) &=&\pm\frac 1{(z-w)}J^{\pm}(w)+...  \nonumber \\
J^3(z)J^3(w) &=&\frac{-k/2}{(z-w)^2}+...  \label{algebra}
\end{eqnarray}
which can be realized in terms of the three fields $\beta, \gamma, \phi$
introduced above. These fields have correlators  given by 
\begin{equation}
<\beta(z)\gamma(w)> = {\frac{1}{z-w}} \quad ; \quad <\phi(z) \phi(w)> = - 
{\rm ln}~(z-w).
\end{equation}
There are also $\bar z$ dependent antiholomorphic fields ($\bar\beta (\bar z)
$, $\bar \gamma (\bar z)$, $\phi(\bar z)$). However we shall discuss the
left moving part of the theory only and assume that all the steps go through
to the right moving part as well, indicating the left-right matching
conditions where necessary.

The $SL(2)$ currents can be thus represented as 
\begin{eqnarray}
J^+(z) &=& \beta  \nonumber \\
J^3(z) &=& -\beta\gamma - {\frac{\alpha_+}{2}}\partial\phi  \nonumber \\
J^-(z) &=& \beta\gamma^2 + \alpha_+ \gamma\partial\phi + k\partial\gamma
\end{eqnarray}
where $\alpha_+ = \sqrt{2(k-2)}$ and $k$ is the level of the $SL(2)$
algebra. They give rise to the Sugawara stress-energy tensor 
\begin{equation}
T_{SL(2)}(z) = \beta\partial\gamma - {\frac{1}{2}}\partial\phi\partial\phi - 
{\frac{1}{\alpha_+}} \partial^2\phi ,  \label{tsl}
\end{equation}
which leads to the following central charge of the Virasoro algebra 
\begin{equation}
c= 3 + \frac{12}{\alpha_+^2} = {\frac{3k}{k-2}}.
\end{equation}

The complete action associated with the energy-momentum tensor (\ref{tsl})
is 
\begin{equation}
S = \int d^2 z \left (\frac 12\partial\phi\bar\partial\phi - \frac
2{\alpha_+}R\phi + \beta\bar\partial\gamma + \bar\beta\partial\bar\gamma
\right ) - S_{int} ,
\end{equation}
where the linear dilaton term can be interpreted as the effect of a
background charge at infinity and $S_{int} = \int d^2 z \beta\bar\beta e^{-2
\phi}$. Notice that when $\phi\rightarrow\infty$ the interaction term
vanishes and the theory can be treated perturbatively. The boundary
of $AdS_3$ is located in this region where the theory can be described
safely in terms of these free fields.

The quantum theory is constructed similarly as Liouville theory where the
interaction term gets renormalized as 
\begin{equation}
{\cal S}_+ = \int d^2 z ~ \beta\bar\beta e^{-{\frac{2}{\alpha_+}}\phi}.
\label{s+}
\end{equation}
This term in the action can be interpreted as a screening charge operator in
correlation functions when the theory is considered as a Coulomb gas model.
In this case there is however another screening operator which is equally
suitable for the purpose of guaranteeing the charge conservation condition
\cite{dotse, ooguri} (see subsection 2.2 below), namely 
\begin{equation}
{\cal S}_- = \int d^2 z ~ (\beta\bar\beta)^{\frac{\alpha_+^2}{2}}
e^{-\alpha_+\phi}.  \label{s-}
\end{equation}
These two perturbations (\ref{s+}) and (\ref{s-}) were discussed in
reference
\cite{andreev} in the framework of the AdS/CFT correspondence, where it
was pointed out that, unlike ${\cal S}_+$, ${\cal S}_-$ cannot reconstruct
the AdS$_3$ geometry
upon integrating out the auxiliary $\beta , \bar \beta$ fields. 
The situation is again similar to Liouville theory where two interaction
terms seem
necessary in order to get the correct pole structure of the correlation
functions \cite{tesch3}. We shall discuss the contribution of both
perturbations in the computation of the two 
and
three point functions in subsection 2.2 below. Here let us notice that
${\cal
S}_+$ can be considered as a small perturbation when $k\rightarrow 2$ for
arbitrary $\phi$ whereas it is finite for $k\rightarrow \infty$.
 The exponential term in ${\cal S}_-$ instead is small when
$k\rightarrow \infty$ but it tends to one for $k\rightarrow 2$.

Vertex operators creating physical states are needed in order to compute
correlation functions. In the next subsection we review the properties of
the
operators that have
been proposed in the literature.

 \subsection{Primary fields and vertex
operators}

The primary fields of the $SL(2)$ conformal theory $\Phi^j_m(z)$ satisfy the
following OPEs with the currents 
\begin{eqnarray}
J^+(z) \Phi^j_m(w) &=& {\frac{(j-m)}{z-w}} \Phi^j_{m+1}(w) + ...  \nonumber
\\
J^3(z) \Phi^j_m(w) &=& {\frac{m}{z-w}} \Phi^j_m(w) + ...  \nonumber \\
J^-(z) \Phi^j_m(w) &=& {\frac{(-j-m)}{z-w}} \Phi^j_{m-1}(w) + ...  \label{pf}
\end{eqnarray}
They can be associated to differentiable functions on $H_3^+$, which can be
decomposed in terms of representations of $SL(2)\times \overline{SL(2)}$.
The differentiable operatos are the zero modes of the algebra. A convenient
plane wave normalizable basis for ${\cal L}^2(H_3^+)$ is given by \cite{ggv,
tesch1} 
\begin{equation}
\Phi_j(z, x) = {\frac{2j+1}{\pi}} \left (|\gamma - x|^2 e^{\phi/\alpha _+}
+
e^{-{\phi/\alpha_+}} \right )^{-2j-2}  \label{func}
\end{equation}
where $(x,\bar x)$ are auxiliary complex variables. Spin $j$ and $-j-1$
representations are equivalent and consequently the functions $\Phi_j(z,x)$
and $\Phi_{-j-1}(z,x)$ satisfy the following relation 
\begin{equation}
\Phi_{j}(z,x) = {\frac{R(j)}{\pi}} \int d^2 y ~ |x-y|^{-4j-4} ~
\Phi_{-j-1}(z,y) ,  \label{relfunc}
\end{equation}
where $R(j)$ is the reflection coefficient verifying 
\begin{equation}
R(j) R(-j-1) = -(2j+1)^2.
\end{equation}

We are interested in the near boundary limit, $i.e.$ $\phi\rightarrow \infty$%
. The expansion of $\Phi_j$ around $\phi \approx \infty$ was worked out in
detail in
references \cite{ks} where it was remarked that the behavior of the
functions (%
\ref{func}) changes as one approaches $j = - 1/2$. In fact, as $%
\phi\rightarrow \infty$, 
\begin{equation}
\lim_{\phi \rightarrow \infty }\Phi _j(x,\bar x)=e^{\frac{2j}{\alpha
_{+}}\phi }\delta ^{(2)}(\gamma -x)+\frac {2j+1}{\pi} \left| \gamma
-x\right| ^{-4j-4}e^{\frac{-2j-2}{\alpha _{+}}\phi }+{\cal O}(e^{\frac
1{\alpha _{+}}j\phi }),  \label{cachabacha}
\end{equation}
and it is easy to see that the first term is leading for $j>-1/2$ whereas
the second one dominates for $j<-1/2$.

Fourier transforming the leading terms in this expansion as 
\begin{equation}
{\cal V}_{j,m,\bar m}(z,\bar z)=\int d^2 x ~ \Phi _j(x,\bar x) ~ x^{j-m} ~
\bar x^{j-\bar m} ,  \label{fourier}
\end{equation}
one obtains 
\begin{equation}
{\cal V}_{j,m,\bar m}= V_{j,m,\bar m} + ~ (2j+1) {\frac{%
\Gamma(j-m+1)\Gamma(j+\bar m +1) \Gamma(-2j-1)}{\Gamma(-j+\bar
m)\Gamma(-j-m) \Gamma(2j+2)}} V_{-1-j,m,\bar m}  \label{vcont}
\end{equation}
where 
\begin{equation}
V_{j,m,\bar m} = \gamma ^{j-m}\bar\gamma^{j-\bar m}e^{\frac{2j}{\alpha _{+}}%
\phi} .
\end{equation}
Notice that equation (\ref{fourier}) is well defined if $m-\bar m \in Z$.
Let us remark that, except for a $j,m$ dependent factor, the field
dependence of
both terms in expression (\ref{vcont}) is related by $j\leftrightarrow
-j-1$.

Consider now the relation (\ref{relfunc}).  Fourier transforming both sides
one finds 
\begin{equation}
{\cal V}_{j,m,\bar m} = R(j) {\frac{\Gamma(j-m+1)\Gamma(j+\bar m +1)
\Gamma(-2j-1)}{\Gamma(-j+\bar m)\Gamma(-j-m) \Gamma(2j+2)}} {\cal V}_{-j-1,
m,\bar m},
\end{equation}
and consequently the relative weight between both terms in the vertex
operator has to be modified as
\begin{equation}
{\cal R}(j,m) = R(j) {\frac{\Gamma(j-m+1) \Gamma(j+\bar m +1) \Gamma(-2j-1)}{%
\Gamma(-j+\bar m)\Gamma(-j-m) \Gamma(2j+2)}} .
\end{equation}
The classical factor $2j+1$ has been replaced by the
reflection coefficient $%
R(j)$ which  was found in reference \cite{tesch1, satoh} to be 
\begin{equation}
R(j) = (2j+1) \left (
\frac{\Gamma(1-\rho)}{\Gamma(1+\rho)} \right
) ^{2j+1} \frac {\Gamma(1+(2j+1)\rho)}{\Gamma\left (1-(2j+1)\rho \right)}
,
\end{equation}
where 
$\rho=-(k-2)^{-1}$. Observe that it
reduces to $2j+1$ as $%
k\rightarrow\infty$. 

Therefore the final form of the vertex operator in the
large $\phi$ limit is 
\begin{equation}
{\cal V}_{j,m,\bar m} =V_{j,m,\bar m} + {\cal R}(j,m) V_{-j-1,m,\bar m}
\label{fullver}
\end{equation}

The first term above reproduces the standard form of the vertex operator
used in the Wakimoto representation of the theory \cite{morozov, becker}.
 This term is dominant in
the large $\phi$ limit for states in the discrete representation
satisfying
the unitarity bound $-1/2<j<(k-3)/2$. However notice that both terms
contribute to the same order for states in  the continuous representation, $%
i.e.$ $j=-1/2 \pm i\lambda$, $\lambda\in R$, $m\in R$, 
 and thus the full expression should be used in this case.

For $j=-1/2$ there is a ``resonance'' and the leading term in the expansion
is 
\begin{equation}
\lim_{\phi \rightarrow \infty }\Phi _{-\frac 12}\sim \frac \phi {\alpha
_{+}}\delta (\gamma -x)\delta (\bar \gamma -\bar x)e^{-\frac {\phi}{\alpha
_{+}}}  \label{laimportancia}
\end{equation}
Observe that the Fourier transform of this expression belongs to the
following functional form
\beq
\hat V_{j,m,\bar m} = \frac
{\phi}{\alpha_+}\gamma^{j-m}\bar\gamma^{j-\bar m} e^{\frac
{2j}{\alpha_+}\phi}
\label{hve}
\eeq
evaluated at $j=-\frac 12$. The operators (\ref{hve})
have logarithmic structure in the Virasoro algebra, namely
\begin{equation}
L_{0}\hat{V}_{j,m}=-\frac{j(j+1)}{k-2}\hat{V}_{j,m}-\frac{1+2j}{k-2}%
V_{j,m} .  \label{mafor}
\end{equation}
However they are indeed primary fields for the particular case $j=-1/2$. 
These objects are called prelogarithmic or puncture operators and
they form a Jordan block of the $SL(2)_k$ Kac-Moody algebra.

\subsection{Two and three point functions}

In this Section we review the computation of two and three point tachyon
amplitudes
performed in the Wakimoto representation of the SL(2,R)/U(1) black hole
background in reference \cite{becker}. We follow closely the steps performed
by K. and M. Becker but we are more general, $i.e.$ we consider the two
interaction terms (\ref{s+}) and (\ref{s-}) in the action. 

Let us start by recalling the calculation of correlation functions of
tachyon vertex operators, namely 
\begin{equation}
{\cal A}^{j_1,j_2,...,j_N}_{m_1,m_2,...,m_N}(z_1,z_2,...,z_N) = \int
\prod_{i=1}^N d^2 z_i \left\langle \prod_{i=1}^N V_{j_i,
m_i}(z_i)\right\rangle .
\end{equation}
The expectation value is taken with the action 
\begin{eqnarray}
S_{\mu ,\tilde{\mu}}[\beta ,\gamma ,\phi ] &=&\frac{1}{4\pi }\int
d^{2}z\left( \partial \phi \bar{\partial}\phi -\frac{2}{\alpha _{+}}R\phi
+\beta \bar{\partial}\gamma +\bar{\beta}\partial \bar{\gamma}\right) + 
\nonumber \\
&&-\frac{1}{4\pi }\int d^{2}z\left( \mu \beta \bar{\beta}e^{-\frac{2}{\alpha
_{+}}\phi }+\tilde{\mu}\beta ^{\frac{\alpha _{+}^{2}}{2}}\bar{\beta}^{\frac{%
\alpha _{+}^{2}}{2}}e^{-\alpha _{+}\phi }\right)  \label{arriba}
\end{eqnarray}
as discussed above, where two coupling constants $\mu$ and $\tilde\mu$ have
been inserted. In the case of the 2D black hole, $\mu$ is related to the
black hole mass (see \cite{dvv}).

Consistently with the perturbative scheme adopted for the computation, the
vertex operators considered correspond to the large $\phi$ limit of the
primary fields for $j>-1/2$ and $m_i=\bar m_i$. 
The
large $\phi$ limit of the vertex operators creating tachyon states in the
coset theory are 
\begin{equation}
V_{j,m,\bar
m}=:\gamma^{j-m}\bar\gamma^{j-\bar m}e^{{\frac{2j}{\alpha_+}}\phi}
e^{im\sqrt{\frac{2}{k}}%
X}: . \label{vcos}
\end{equation}
The compact
free boson $X$ was introduced in \cite{dvv} to gauge the U(1) subgroup. 
The integration over its zero mode yields the following conservation
law 
\begin{equation}
\sum_{i=1}^N m_i = 0.  \label{clm}
\end{equation}
Other than that, the contribution of $X$ completes the conformal properties
of the N-point functions, therefore we  omit explicit reference to 
this field in the correlators.

The functional integral over $\phi$ can be performed as usual splitting the
field $\phi$ into zero mode and oscillator parts, $\phi(z) = \phi_0 + \tilde
\phi(z)$. Using the Gauss-Bonnet theorem and performing some algebraic
manipulations, we find on the sphere 
\[
\left\langle \prod_{n=1}^{N}V_{j_{n},m_{n},\bar{m}_{n}}\right\rangle _{\mu ,%
\tilde{\mu}}=\Gamma \left( -s_{-}\right) \Gamma \left(-
s_{+}\right) \mu ^{s_{+}}\tilde{\mu}^{s_{-}}\times 
\]
\begin{eqnarray}
&&\times \left\langle \prod_{n=1}^{N}V_{j_{n},m_{n},\bar{m}%
_{n}}\prod_{r=1}^{s_{+}}{\cal S}_{+}(w_{r})\prod_{t=1}^{s_{-}}{\cal S}%
_{-}(y_{t})\right\rangle _{\mu =0,\tilde{\mu}=0} \delta \left( s_{-}\frac{%
\alpha _{+}^{2}}{2}+s_{+}-1-\sum_{n=1}^{N}j_{n}\right)
\end{eqnarray}
where we have absorbed an overall constant in the definition of the path
integral.
This expression shows that the interaction terms play the role of screening
operators 
\begin{equation}
{\cal S}_+ = \int d^2 y ~ \beta(y) \bar \beta(\bar y) e^{-{\frac{2}{\alpha_+}%
}\phi(y,\bar y)}
\end{equation}
and 
\begin{equation}
{\cal S}_- = \int d^2 w ~ \beta(w)^{\frac{\alpha_+^2}{2}} \bar \beta(\bar
w)^{\frac{\alpha_+^2}{2}} e^{-\alpha_+\phi(w,\bar w)},
\end{equation}
which have to be introduced in order to satisfy the conservation law 
\begin{equation}
(k-2) s_-+s_+ =\sum_{i=1}^N j_i +1 ,  \label{cls}
\end{equation}
arising from the integration over $\phi_0$. Notice that $s_-$ and $s_+$ are
in general non integer, and they can be complex numbers if the states belong
to the continuous representation. Without loss of generality we shall work
out in full detail the case $s_+=0$, contrary to the choice $s_-=0$
performed in reference \cite{becker}. At the end we shall compare the
results obtained in both cases.

Let us start by considering two point functions of tachyon vertex operators.
Conformal invariance allows one to determine the general structure of the
two point functions, namely 
\begin{equation}
< V_{j_1,m_1}(z_1) V_{j_2,m_2}(z_2)> = |z_1-z_2|^{-4\Delta_1} \left [
A(j_1)\delta(j_1+j_2+1) + B(j_1)\delta(j_1-j_2)\right ]  \label{2pf}
\end{equation}
where $\Delta_1 = j_1(j_1+1)\rho + m_1^2/k$ is the conformal weight of
the vertex operators (\ref{vcos}) in the coset theory.

The correlators to be computed in order to determine $A(j)$ and $B(j)$ are 
\begin{equation}
\int \prod_{i=1}^{s_-} d^2 w_i
\left\langle \gamma _{(z_1)}^{j_1-m_1}\gamma
_{(z_2)}^{j_2-m_2}\prod_{i=1}^{s_-}\beta^{k-2} _{(w_i)}\right\rangle
\times
c.c. \times \left\langle e^{\frac{2j_1}{\alpha _{+}}\phi (z_1,\bar
z_1)}e^{%
\frac{2j_2}{\alpha _{+}}\phi (z_2,\bar z_2)}\prod_{i=1}^{s_-}e^{-{\alpha
_{+}}%
\phi (w_i,\bar w_i)}\right\rangle ,  \label{2ptf}
\end{equation}
where we have renamed $\tilde\phi = \phi$, and the conservation law
(\ref{cls}%
) to be considered in this case is 
\begin{equation}
s_- = -\rho(j_1+j_2+1).  \label{s2pf}
\end{equation}

The $\beta-\gamma$ correlator can be computed by bosonization, introducing
as usual two ordinary bosons $u$ and $v$ such that
\begin{equation}
\beta = -i\partial v e^{iv-u} \quad , \quad \gamma = e^{u-iv} ,
\end{equation}
where 
\begin{equation}
<u(z)u(w)> = <v(z)v(w)> = - {\rm ln}(z-w).
\end{equation}
This allows to make sense of non positive integer powers of $\gamma$,
whereas non integer numbers of $\beta$ fields are not well defined. 
Technical difficulties arising from the occurrence of fractional powers
of $\beta - \gamma$ fields have been dealt with in references \cite{bega}.
 However
we shall proceed as if there were an integer number of screening operators
with an integer power of $\beta$ fields and at the end the condition (\ref
{s2pf}) for $s_-$ will be imposed, assuming the expressions are defined by
analytic continuation in $s_-$. The agreement of the results with those
obtained by other approaches supplies the justification for this
procedure.

Clearly the computation of the first term $A(j)$ in (\ref{2pf}) requires no
screening operators. Contractions from the $\gamma^{\prime}$s of the vertex
operators are equal to one and the contribution from the exponentials
reproduce the overall factor $|z_1-z_2|^{-4\Delta_1}$. Therefore, $A(j)=1$.

The term $B(j)$ in (\ref{2pf}) can be
computed similarly as in reference \cite{becker}. It is convenient to
fix the tachyon vertices at $z_1 = 0$ and $z_2 = 1$ and the
position of one screening operator at $w_{s_-} = \infty$ in order to factor
out the SL(2,C) invariant volume. 

The contribution of the $\beta - \gamma$ system can be obtained generalizing
the procedure in reference \cite{becker}, where it was found that 
\begin{equation}
\left\langle \gamma _{(0)}^{j_1-m_1}\gamma
_{(1)}^{j_2-m_2}\prod_{i=1}^{s}\beta _{(w_i)}\right\rangle = {\cal P}^{-1} 
\frac{\partial^s {\cal P}}{\partial w_1 ... \partial w_s}
\end{equation}
with 
\begin{equation}
{\cal P} = \prod_{i=1}^s w_i^{m_1-j_1} (1-w_i)^{m_2-j_2}\prod_{i<j}(w_i-w_j)
\end{equation}

In order to compute the correlator in equation (\ref{2ptf}) it is convenient
to point split the insertion points of the screening operators, $i.e.$ take $%
(k-2)s_-$ different  points as $w_r^{(n)}, r=1,...,s; n=
1,...,k-2$ and at the
end take the limit where $w_{r}^{(n)}\rightarrow w_r$, $\forall n$. Thus we
obtain

\begin{equation}
\left\langle \gamma _{(0)}^{j_1-m_1}\gamma
_{(1)}^{j_2-m_2}\prod_{i=1}^{s_-}\beta^{k-2} _{(w_i)}\right\rangle =
\lim_{w_i^{(n)}\rightarrow w_i^{(1)}=w_i} {\cal P}^{-1} \frac{%
\partial^{(k-2)s_-}{\cal P}}{\partial w_1^{(1)}...\partial w_1^{(k-2)} ...
\partial w_{s_-}^{(1)} ... \partial w_{s_-}^{(k-2)}}
\label{pes}
\end{equation}
with 
\begin{equation}
{\cal P} = \prod_{i=1}^{s_-} \prod _{n=1}^{k-2}
(w_i^{(n)})^{m_1-j_1}(1-w_i^{(n)})^{m_2-j_2}\prod_{i<j}(w_i^{(n)}-w_j^{(m)}).
\end{equation}

Therefore the contribution from the $\beta - \gamma$ correlator is 
\begin{equation}
\left |\left\langle \gamma _{(0)}^{j_1-m_1}\gamma
_{(1)}^{j_2-m_2}\prod_{i=1}^{s_-}\beta^{k-2} _{(w_i)}\right\rangle \right
|^2 = (-)^{-\tilde\rho s_-} \triangle(1+j_1-m_1)\triangle(1+j_2-m_2)
\prod_{i=1}^{s_-} |w_i|^{2\tilde\rho}|1-w_i|^{2\tilde\rho},
\end{equation}
where $\Delta(x) = \Gamma(x)/\Gamma(1-x)$,
$\tilde\rho=\rho^{-1}=-(k-2)$ and $m_i = \bar m_i$.

Performing the $\phi$ contractions, the $(s_--1)$ integrals from the
screening operators are 
\begin{equation}
\int \prod_{i=1}^{s_--1} d^2 w_i |w_i|^{4j_1+2\tilde\rho}
|1-w_i|^{4j_2+2\tilde\rho} \prod_{i<j}|w_i-w_j|^{4\tilde\rho}.
\label{albe}
\end{equation}
These integrals have been computed by Dotsenko and Fateev \cite{df} who
found 
\begin{eqnarray}
\int \prod_{i=1}^s d^2 w_i \prod _{i=1}^s
|w_i|^{2\alpha}|1-w_i|^{2\beta} \prod_{i<j}^s |w_i-w_j|^{4\sigma} =
s!\pi^s
\left (\triangle(1-\sigma)\right )^s \prod_{i=1}^s
\triangle(i\sigma)\times 
\nonumber \\
\times \prod_{i=0}^{s-1}\triangle(1+\alpha+i\sigma)
\triangle(1+\beta+i\sigma) \triangle(-1-\alpha-\beta-(s-1+i)\sigma).
\label{dotf}
\end{eqnarray}

Specifying the particular values of $\alpha$ and $\beta$ in (\ref{albe}),
the
final result obtained for the term $B(j)$ in the two-point
functions is 
\begin{equation}
B(j) = (-\tilde\mu\pi \triangle(-\tilde\rho))^{s_-}
\triangle(1+j-m)\triangle(1+j+m)s_-\tilde\rho^2
\triangle(1-s_-)\triangle(\tilde\rho s_-) \delta(m_1+m_2),  \label{wwww}
\end{equation}
where $j=j_1=j_2$ and $m=m_1=-m_2=\bar m$.

Let us first compare this expression with the result obtained in reference 
\cite{becker}, where only screening operators ${\cal S}_+$ were considered,
namely \footnote{%
In comparison to \cite{becker} it should be noted that the result here
contains an extra factor $\rho^2$}

\begin{equation}
B(j) = (-\pi\mu \triangle(-\rho))^{s_+} \triangle(1+j-m)\triangle(1+j+m)
s_+\rho^2\triangle(1-s_+)\triangle(\rho s_+) \delta(m_1+m_2), 
\label{becker}
\end{equation}
where $s_+=j_1+j_2+1$.

The situation is similar to the case of the minimal models where there are
two screening charges as well. The conformal properties of the correlation
functions cannot be changed by the insertion of ${\cal S}_{\pm}$, as long as
the correlators satisfy the charge balance. Therefore  the results (\ref
{wwww}) and (\ref{becker}) should coincide, $i.e.$ they should be
independent of the screening operator used.

In order to see if this is the case  let us recall that, as was shown
in \cite{becker}, the
two-point functions can be obtained from the three-point functions
containing one highest weight state $V_{j_1,j_1}$, taking the limit $j_1=
i\varepsilon\rightarrow 0$. We shall show below, after computing the three
point functions, that in this case the term $B(j)$ is as found above ($i.e.$
equations (\ref{wwww}) and (\ref{becker}) from the direct calculation) with
an extra factor $(s_-\tilde\rho)^{-1}$ and $(s_+\rho)^{-1}$, respectively.
Thus we shall compare the following expressions 
\begin{equation}
B_-(j) = (-\pi\tilde\mu \triangle(-\tilde\rho))^{s_-}
\triangle(1+j-m)\triangle(1+j+m)\tilde\rho
\triangle(1-s_-)\triangle(\tilde\rho s_-) \delta(m_1+m_2).
\label{b-}
\end{equation}
and 
\begin{equation}
B_+(j) = (-\pi\mu \triangle(-\rho))^{s_+}
\triangle(1+j-m)\triangle(1+j+m)\rho\triangle(1-s_+)\triangle(\rho s_+)
\delta(m_1+m_2).  \label{b+}
\end{equation}

It is interesting to notice that these two expressions agree when
replacing $%
s_+=j_1+j_2+1$ and $s_-=-\rho(j_1+j_2+1)$, if the following expression holds 

\begin{equation}
\pi\tilde\mu\triangle(-\tilde\rho) = (\pi\mu\triangle(-\rho))^{-\rho^{-1}}.
\label{liute}
\end{equation}

We shall find a similar relation in the computation of three point
functions, after which we postpone some comments.

Let us now compare the results obtained above with others found in the
literature. The expression for $B_+(j)$ (\ref
{b+}) is exactly the Fourier transform of the result obtained in
references \cite{teschner, satoh} (see \cite{gk}), namely 
\begin{equation}
\left\langle \Phi _{j_{1}}(x,\bar{x})\Phi _{j_{2}}(x^{\prime },\bar{x}%
^{\prime })\right\rangle =\frac{k-2}{\pi }\left[ \nu (k)\right]
^{j_{1}+j_{2}+1}\frac{\Gamma (1-\frac{j_{1}+j_{2}+1}{k-2})}{\Gamma (\frac{%
j_{1}+j_{2}+1}{k-2})}|x-x^{\prime }|^{-2(j_{1}+j_{2}+2)}
\end{equation}
where 
\begin{equation}
\nu(k) = \frac 1\pi \frac{\Gamma(1-\rho)}{\Gamma(1+\rho)},
\end{equation}
except for an irrelevant factor $(\rho/\pi^2 )^{s_+}$ (notice that when
$%
s_+=0$ this factor is 1, thus this does not affect the term $A(j)$ in the
2-point functions). Moreover, the Fourier transform 
can be performed even when $j_1+j_2+1=0$.
In order to see that, recall that the Dirac delta function can
be written as 
\[
\delta ^{(2)}(x)=\frac{1}{\pi }\lim_{\varepsilon \rightarrow 0}\frac{%
|x|^{2(\varepsilon -1)}}{\Gamma (\varepsilon )}.
\]
Then defining $j_{2}=-1-j_{1}-\varepsilon $ and taking the limit $%
\varepsilon \rightarrow 0$, the term $A(j)$ ($i.e.$
the term which is proportional to $\delta (j_{1}+j_{2}+1))$ 
is recovered, namely
\[
\lim_{\varepsilon \rightarrow 0}\left\langle \Phi _{j_{1}}(x,\bar{x})\Phi
_{-1-j_{1}-\varepsilon }(x^{\prime },\bar{x}^{\prime })\right\rangle
=\delta^2 (x-x^{\prime }).
\]

It is remarkable that the free field approximation reproduces so accurately
the exact result. 
We shall now show that the same agreement is found for the three
point
functions.

The computation of the three tachyon amplitudes goes along the same steps.
It
turns out that the simplest way to do it is to start with one highest weight
tachyon, for example take $j_2=m_2$. It was shown in
reference \cite{becker} that a general three tachyon amplitude can
then be expressed as a function of this one, acting with the currents
$J^-$.

It is convenient to fix the positions of the vertices at
$(z_{1},z_{2},z_{3})=(0,1,%
\infty )$. The amplitude is then
\begin{eqnarray}
{\cal A}^{j_{1},j_{2},j_{3}}_{m_1,m_2,m_3} &=& \Gamma(-s_-)\int
\prod_{i=1}^{s_{-}}d^{2}w_{i}\left\langle \gamma
_{(0)}^{j_{1}-m_{1}}\gamma
_{(\infty )}^{j_{3}-m_{3}}\prod_{i=1}^{s_{-}}\beta
^{k-2}_{(w_{i})}\right\rangle
\times c.c.\times   \nonumber \\
&&\times \left\langle e^{\frac{2j_{1}}{\alpha _{+}}\phi (0)}e^{\frac{2j_{2}}{%
\alpha _{+}}\phi (1)}e^{\frac{2j_{3}}{\alpha _{+}}\phi (\infty
)}\prod_{i=1}^{s_-}e^{-{\alpha _{+}}\phi (w_{i},\bar{w}%
_{i})}\right\rangle 
\end{eqnarray}
and the conservation laws are in this case
\beq
(k-2) s_-=j_1+j_2+j_3+1 \quad , \quad m_1+m_2+m_3=0
\eeq
(notice that we are again using screening operators ${\cal S}_-$, unlike
reference \cite{becker}).

The $\beta - \gamma$ correlator can be evaluated as in equation
(\ref{pes}) above, with
\beq
{\cal P} = \prod_{i=1}^{s_-}\prod_{n=1}^{k-2}
(w_i^{(n)})^{m_1-j_1}\prod_{i<j} (w_i^{(n)}-w_j^{(m)}),
\eeq
and the result is now
\beq
\left |\left\langle
\gamma^{j_1-m_1}_{(0)}\gamma^{j_3-m_3}_{( \infty )}\prod_{i=1}^{s_-}
\beta^{k-2}_{(w_i)}\right\rangle \right | ^2
= (-)^{-\tilde\rho s_-}\triangle(1+j_1-m_1) \triangle
(1+j_3-m_3)\prod_{i=1}^{s_-}|w_i|^{2\tilde\rho}
\eeq

 After performing the contractions of the exponentials we obtain 
\begin{equation}
{\cal A}^{j_1,j_2,j_3}_{m_1,j_2,m_3} = (-)^{-\tilde\rho s_-}
\triangle(1+j_1-m_1)\triangle(1+j_3-m_3) \tilde{\cal I}(j_1,j_2,j_3,k),
\label{3pf}
\end{equation}
where 
\begin{equation}
\tilde{\cal I}(j_1,j_2,j_3,k) =
\tilde\mu^{s_-}\Gamma(-s_-)\int\prod_{i=1}^{s_-}
d^2w_i
|w_i|^{4j_1 +2\tilde\rho} |1-w_i|^{4 j_2} \prod_{i<j}
|w_i-w_j|^{4\tilde\rho}
\end{equation}
can be evaluated using the Dotsenko Fateev formula (\ref{dotf}) and the
special function $G(j)$ which satisfies the following relations
\ber
G(j) = G(-1-j-(k-2)), \nonumber\\
G(j+1) = \triangle ((1+j)\rho) G(j), \nonumber\\
G(j-k+2) = (k-2)^{-2j-1} \triangle (1+j) G(j).
\label{fung}
\eer

The final result can be expressed as 
\begin{equation}
\tilde{\cal I}(j_1,j_2,j_3,k) = \frac{1}{k-2}
[\pi\tilde\mu\triangle(-\tilde\rho)]^{s_-} D(j_1,j_2,j_3)
\end{equation}
where 
\beq
 D(j_1,j_2,j_3) = \triangle (-1-\sum_i j_i)
\triangle(1+2j_2)\triangle(j_1-j_3-j_2) \triangle(-j_1-j_2+j_3)
C(j_1,j_2,j_3)
\eeq
and 
\ber
C(j_1,j_2,j_3) =
\frac{G(-2-\sum_i
j_i)}{G(-1)}
\frac{G(-j_1+j_2-j_3-1)}{G(-2j_1-1)}\times
\nonumber\\
\times \frac{ G(j_1-j_3-j_2-1) G(-1-j_1-j_2+j_3)}
{G(-2j_3-1)G(-2j_2-1)} .
\label{tildedd}
\eer

This computation  can be repeated using the screening operators
${\cal
S}_+$, as it was done in reference \cite{becker},
 instead of ${\cal S}_-$. 
The Dotsenko Fateev integral is in this case
\beq
{\cal I}(j_1,j_2,j_3,k) = \mu^{s_+}\Gamma(-s_+)\int\prod_{i=1}^{s_+}d^2
w_i |w_i|^{-2-4\rho j_1} |1-w_i|^{-4\rho j_2} \prod_{i<j}|w_i-w_j|^{4\rho}
\label{3pfs+}
\eeq
Taking into account the multiplicity factor from the $\beta -\gamma$
system, the final result is 
\begin{equation}
{\cal A}^{j_1,j_2,j_3}_{m_1,j_2,m_3} = (-)^{-s_+}
\triangle(1+j_1-m_1)\triangle(1+j_3-m_3){\cal I}(j_1,j_2,j_3,k),
\label{3pf+}
\eeq
where
\beq
{\cal I} (j_1,j_2,j_3, k) = (k-2)
[\pi\mu\triangle(-\rho)]^{s_+} D(j_1,j_2,j_3).
\eeq

Therefore here again, similarly as in the computation of the two point
functions, both results (\ref{3pf}) and (\ref{3pf+}) are related by
an exchange of $\rho\leftrightarrow \tilde\rho$ and $s_+\leftrightarrow s_-$.

It is easy to see that (\ref{3pf+}) completely agrees
with the 
 Fourier transform of the result obtained in
reference \cite{tesch1}, namely 
\begin{eqnarray}
\int d^2 x_1 d^2 x_2 d^2 x_3 |x_1|^{2(j_1-m_1)} |x_2|^{2(j_2-m_2)}
|x_3|^{2(j_3-m_3)} |x_1-x_2|^{-2(j_1+j_2-j_3+1)}\times  \nonumber \\
|x_1-x_3|^{-2(j_1+j_3-j_2+1)} |x_2-x_3|^{-2(j_2+j_3-j_1+1)}
C(j_1,j_2,j_3),
\end{eqnarray}
which can be explicitly performed in the case $j_2=m_2$.
The pole structure of this expression was analysed
in reference \cite{gk}.

Let us now make some comments about equation (\ref{liute}).
A similar relation is discussed in reference \cite{tesch3} in connection
with Liouville theory, where it
reflects the self duality of the theory when
$\rho\leftrightarrow\tilde\rho$ and moreover it seems necessary to produce 
the correct pole structure of the correlators. However the SL(2)/U(1)
coset theory is not obviously self dual. Indeed it has been 
 conjectured to be equivalent to the
Sine-Liouville model, $i.e.$ $c=1$ CFT coupled to a Liouville field
\cite{zamo, kkk} (see reference \cite{hori} for the fermionic
generalization of this duality). 
In this case there is a strong/weak coupling duality on
the world
sheet. The cigar CFT becomes weakly coupled in the limit $k\rightarrow
\infty$ whereas it is strongly coupled in the limit $k\rightarrow 2$,
where the Sine-Liouville theory becomes weakly coupled. Recalling the
comment
below equations (\ref{s+}) and (\ref{s-}), the screening
operators can be observed to satisfy a similar relation, $i.e.$
${\cal S}_-$ is weakly coupled when $k\rightarrow \infty$ and strongly
coupled when $k\rightarrow 2$, contrary to ${\cal S}_+$.
Furthermore both perturbations satisfy a relation of the same sort as
the corresponding interaction terms in
Liouville theory, namely ${\cal S}_- = ({\cal S_+})^{-1/\rho}$.
Similarly, the coupling constants can be seen to be related by
$\tilde\mu\sim \mu^{k-2}$ under a field redefinition.

It is not clear to us what conclusion can be drawn from these
observations. Expressions (\ref{b-})
and (\ref{b+}) might describe the two point functions in two different
regimes of the same theory (similarly (\ref{3pf}) and (\ref{3pf+}) for
the three
point functions). Since the calculations are
perturbative one would expect that (\ref{b-}) and (\ref{3pf}) provide the
correct answer when
$k\rightarrow\infty$ and (\ref{b+}) and (\ref{3pf+}) when $k\rightarrow
2$. However if the
identity (\ref{liute}) is taken seriouly it might be indicating 
 a hidden self duality of the SL(2,R)/U(1) coset theory.

For the sake of completeness we end this Section with a derivation of
equation (\ref{b+}), $i.e.$ the two-point function obtained from the three
point function containing one highest weight state, $j_2=m_2$ in the limit
$j_2=i\varepsilon \rightarrow 0$.  Considering screening operators ${\cal
S}_+$ we obtain 
\beq
{\cal A}^{j_1,i\varepsilon, j_3}_ {m_1,i\varepsilon,m_3} = (-)^{s_+}
\triangle (1+j_1-m_1) \triangle(1+j_3-m_3) I(j_1,j_3,k)
\eeq
where $I(j_1,j_3,k)$ is given by (\ref{3pfs+}) with $j_2=i\varepsilon$,
$i.e.$
simplifying the products of $\triangle$-functions we obtain
\ber
I(j_1,j_3,k)= (\pi\mu\triangle(-\rho))^{s_+}
\triangle(1-s_+)\triangle(\rho s_+)
 \lim_{\varepsilon \rightarrow
0}\triangle(1-2\rho\varepsilon i)\triangle((\varepsilon i
-j_1+j_3)\rho)\times \nonumber\\
\triangle((\varepsilon i + j_1-j_3)\rho).
\eer
The limit $\varepsilon \rightarrow 0$ can be evaluated using that in this
region 
\beq
 \Gamma(-n+\varepsilon) =
\frac{(-)^n}{\varepsilon \Gamma(n+1)} + {\cal O}(1) \quad {\rm for}\quad 
n\in N
,
\eeq
and taking into account the following representation of the delta function
\beq
\delta(j_1-j_3) = \lim_{\varepsilon \rightarrow 0} \frac 1\pi
\frac{\varepsilon}{\varepsilon ^2 + (j_1-j_3)^2}.
\eeq
Putting all this together we obtain $(2\pi i)$ times equation (\ref{b+}).

 \section{The spectral flow and new representations}

Until now we have been considering the SL(2,R)/U(1) WZW model. In this
Section
 we extend the computations of the previous one to string theory in
AdS$_3$.
As it was pointed out in references \cite{hwang, mo}, the
algebra (\ref{algebra}) has a spectral flow symmetry given by 
\begin{eqnarray}
J_n^3 &\rightarrow &\tilde J_n^3=J_n^3-\frac k2\omega \delta _{n,0} 
\nonumber \\
J_n^{\pm } &\rightarrow &\tilde J_n^{\pm }=J_{n\pm \omega }^{\pm }
\label{spectralflow}
\end{eqnarray}
and thus, 
\begin{equation}
L_n\rightarrow \tilde L_n=L_n+\omega J_n^3-\frac k4\omega ^2\delta _{n,0}
\label{sf}
\end{equation}
where $\omega \in Z$ is the winding number. The spectral flow generates new
representations of the $SL(2,R)$ algebra. The Hilbert space of string theory
in AdS$_3$ can be consequently extended ${\cal H}\rightarrow {\cal H}_\omega 
$ in order to include the states $\left| \tilde j,\tilde m,
\omega\right\rangle $ obtained by spectral flow, which satisfy the following
on-shell condition

\begin{equation}
(L_0-1)\left| \tilde j,\tilde m, \omega\right\rangle =\left( -\frac{\tilde
j(\tilde j+1)}{(k-2)}-\omega \tilde m-\frac k4\omega ^2+{\cal N}-1\right)
\left| \tilde j,\tilde m, \omega\right\rangle =0  \label{universall}
\end{equation}
${\cal N}$ being the excitation level of the string. The new representations
are denoted by $\hat{{\cal D}}_{\tilde j}^{\pm,\omega}$ and ${\hat{{\cal C}}%
^\omega_{\tilde j}}$ and they consist of the spectral flow of the discrete
(highest and lowest weight) and continuous series respectively. It was shown
in reference \cite{mo} that the spectrum of the free theory is closed
under the spectral flow
symmetry if the spin $\tilde j$ of the physical states in the discrete
representations is restricted to $\tilde j <{\frac{k-3 }{2}}$.

The spectrum of string theory consists then  of a product of left and right
representations $\hat{{\cal C}}^\omega_{\tilde j, L}\times \hat{{\cal C}}%
^\omega_{\tilde j, R}$ and $\hat{{\cal D}}_{\tilde j, L}^{\pm,\omega} \times 
\hat{{\cal D}}_{\tilde j, R}^{\pm,\omega}$ with the same amount of spectral
flow and the same spin $\tilde j$ on the holomorphic and antiholomorphic
parts and with $-1/2<\tilde j<(k-3)/2$. The partition function containing
the spectral flow of the discrete representations with this bound on the
spin $\tilde j$ was shown to be modular invariant in \cite{mo}. Moreover,
the partition function for thermal AdS$_3$ backgrounds was also found to be
modular invariant and consistent with this spectrum in \cite{mos}. From now
on we drop the tilde on $\tilde j, \tilde m$.

In reference \cite{gn2} we considered this theory as the coset model $\frac{%
H^+_3}{U(1)}\times time$. A direct extension of Dotsenko's method to compute
the conformal blocks in the compact SU(2) CFT to the non-compact $H^+_3$
group manifold was found adequate to deal with the spectral flow symmetry in
vertex operators and scattering amplitudes. Two free scalar fields were
introduced: $X(z)$ gauges the U(1) subgroup as before and the timelike
scalar field $%
Y(z)$ bosonizes the $J^3$ current as 
\begin{equation}
J^3(z) = -i \sqrt{\frac{k}{2}} \partial Y(z) .
\end{equation}
Their propagators are $<Y(z) Y(w)> = -<X(z) X(w)> = {\rm ln} (z-w)$.

In terms of Wakimoto free fields the vertex operators in the unflowed sector
of the theory can be written in the form
\begin{equation}
V_{j,m,\bar m} = \gamma^{j-m}\bar\gamma^{j-\bar m}
e^{{\frac{2j}{\alpha_+}}\phi} e^{i\sqrt{\frac{2}{k}}m
X} e^{i\sqrt{\frac{2}{k}}m Y}.
\end{equation}

Taking into account the spectral flow, for every field $V_{j,m}$ in the
sector $\omega=0$ one can write a field in the sector twisted by $\omega$ as 
\begin{equation}
V^\omega_{j,m} = \gamma^{j-m}
\bar\gamma^{j-\bar m} e^{{\frac{2j}{\alpha_+}}\phi} e^{i\sqrt{\frac{2%
}{k}}m X} e^{i\sqrt{\frac{2}{k}} (m +\omega k/2)Y}.  \label{mov}
\end{equation}
This vertex operator has the following conformal weight 
\begin{equation}
\Delta(V^\omega_{j,m}) = -{\frac{j(j+1)}{k-2}} - m\omega - {\frac{k\omega^2}{%
4}}
\end{equation}
(see reference \cite{argurio} for an alternative approach to the
description of
winding strings).

The $N$-point functions were constructed in \cite{gn2} (the reader is
referred to that reference for details of the construction). They take the
form 
\begin{equation}
{\cal A}_N^{0,\pm} = <\prod_{i=1}^{N-1}V_{j_i,m_i}^{\omega_i}(z_i) \tilde
V_{j_N, j_N}^{\omega_N (0),(\pm)}(z_N) \prod_{n=1}^s {\cal S}(u_n)>_{0,\pm},
\label{fc}
\end{equation}
where the conjugate highest weight operators $\tilde V^{\omega
(0),(\pm)}_{j,j}$ are needed to avoid
redundant integrations. Two of these conjugate vertices were
found to be 
\begin{equation}
\tilde V^{\omega (0)}_{j,j} = \beta^{2j+k-1}
\bar\beta^{2j+k-1} e^{-{\frac{2(j-1+k)}{\alpha_+}}%
\phi}e^{i\sqrt {\frac{2}{k}}j X} e^{i\sqrt{\frac{2}{k}}(j+{\frac{k}{2}}%
\omega)Y}  \label{fcon0}
\end{equation}
and 
\begin{equation}
\tilde V^{\omega (-)}_{j,j} = \beta^{2j} \bar\beta^{2j} 
e^{-{\frac{(2j+k)}{\alpha_+}}%
\phi}e^{i\sqrt {\frac{2}{k}}(j-{\frac{k}{2}}) X}
e^{i\sqrt{\frac{2}{k}}(j+{%
\frac{k}{2}}\omega)Y} . \label{fcon-}
\end{equation}
The screening operators ${\cal S}$ in (\ref{fc}) can be either ${\cal
S}_+$ or ${\cal S}_-$ or combinations of them, as discussed in Section 2.
Similarly to the SU(2) case the conjugate vertices have to be included in
the conformal blocks in the
Coulomb gas formalism to avoid redundant contour
integrations. We now briefly review the  procedure to be followed in order
to find them. 

Non-vanishing correlators must satisfy the charge asymmetry
conditions which are determined by the operator conjugate to the identity.
This is an operator that commutes with the currents and has zero conformal
dimension. Three such operators were found in \cite{gn2} 
\begin{equation}
\tilde {{\cal I}}_0 (z) = \beta^{k-1} e^{{\frac{2(1-k)}{\alpha_+}}\phi}
\quad , \quad \tilde{{\cal I}}_+ = \gamma^{-k} e^{-{\frac{k}{\alpha_+}}\phi}
e^{i\sqrt{\frac{k}{2}}X} \quad ; \quad \tilde{{\cal I}}_- = e^{-{\frac{k}{%
\alpha_+}}\phi} e^{-i\sqrt{\frac{k}{2}}X}
\end{equation}
They lead respectively to the following charge asymmetry conditions 
\begin{equation}
{\cal C}^{(0)} : \quad N_{\beta} - N_{\gamma} = k-1 \quad , \quad \sum_i
\alpha_i = {\frac{2-2k }{\alpha_+}} \quad , \quad \sum_i \xi_i = 0,
\label{carga0}
\end{equation}

\begin{equation}
{\cal C}^{(+)} : \quad N_{\beta} - N_{\gamma} = k \quad , \quad \sum_i
\alpha_i = - {\frac{k}{\alpha_+}} \quad , \quad \sqrt{\frac{2}{k}} \sum_i
\xi_i = \sqrt{\frac{k}{2}},  \label{carga+}
\end{equation}
and

\begin{equation}
{\cal C}^{(-)} : \quad N_{\beta} - N_{\gamma} = 0 \quad , \quad \sum_i
\alpha_i = - {\frac{k}{\alpha_+}} \quad , \quad \sqrt{\frac{2}{k}} \sum_i
\xi_i = - \sqrt{\frac{k}{2}},  \label{carga-}
\end{equation}
where $N_\beta, N_\gamma$ refer to the number of $\beta, \gamma$ fields, $%
\alpha_i$ denotes the coefficient of the field $\phi(z_i)$ in the
exponentials and $\xi_i$ denotes the $``charge"$ of the field $X(z_i)$.
These have to be supplemented with  an additional charge conservation law
arising from exponentials of the field $Y(z)$, namely 
\begin{equation}
\sum_i \Omega_i = \sum_i \left (m_i + \frac{\omega_i k}{2} \right )= 0
\label{win}
\end{equation}
where $\Omega_i$ denotes the $``charge"$ of the field $Y(z_i)$. This is the
energy conservation condition.

The conjugate representations for the highest weight operators (\ref{fcon0})
and (\ref{fcon-}) are found by asking that the two point functions
$<V_{j,j}
\tilde V_{j,-j}>$ do not require screening operators to satisfy the charge
asymmetry conditions ${\cal C}^{(0)}$ and ${\cal C}^{(-)}$ respectively, and
it is easy to see that the conjugate operators $\tilde {V}_{j,j}^{\omega (+)}
$ with respect to ${\cal C}^{(+)}$ do not have such a simple form.

Notice that a highest weight state is included in the definition of the
correlation functions (\ref{fc}). This is done for the sake of simplicity.
More generally other vertex operators can be defined by acting on the
highest weight ones with the currents $J^{-}$, although in practice
more complicated expressions are generated in this way.  Hence the
$N$-point
functions to be considered in the coset theory take the form 
\begin{equation}
{\cal A}_N^{0,-} = < \prod_{i=1}^{N-1}{V}^{\omega_i}_{j_i,m_i} (z_i)
\tilde
{V}_{j_N,j_N}^{\omega_N (0),(-)}(z_N) \prod_{n=1}^{s_+} {\cal S}%
_+(u_n)\prod_{m=1}^{s_-} {\cal S}_-(v_m)>_{0,-},  \label{corr}
\end{equation}
where the number of screening operators $s_+$, $s_-$, must satisfy the
charge asymmetry
conditions (\ref{carga0}) or (\ref{carga-}) respectively,  which 
are determined by the conjugate vacuum state, and
non-vanishing results require in addition the conservation law 
(\ref{win}).

Let us stress that it is possible to construct correlators violating
winding number
conservation by, for instance, inserting conjugate operators $\tilde
V^{\omega (-)}_{j,j}$ instead of direct ones into ${\cal A}_N^{(0)(-)}$.
In
fact, correlation functions containing $K$ of these conjugate operators lead
to $\sum_i \omega_i = -K$ when combining (\ref{carga0}) or (\ref{carga-}) 
with (\ref{win}),
whereas processes conserving winding number ($\sum_i\omega_i=0$) are
obtained
when inserting direct vertex operators. Recall that it is possible to
consider correlators containing up to $N-2$ conjugate operators of a
different kind as the one which is required by the conjugate vacuum state,
and thus the
winding number conservation can be violated up to $N-2$ units (this
possibility was proposed in \cite{zamo, gk}).

Let us now proceed to compute two and three point functions using this
formalism.

The general structure of the two point functions is determined from
conformal invariance to be as equation (\ref{2pf}) above, 
namely
\beq
\left\langle V_{j_1,m_1}^{\omega_1}(z)
V_{j_2,m_2}^{\omega_2}(w)\right\rangle
= |z-w|^{-4\Delta_1} [A^\omega(j_1) \delta(j_1-j_2) + B^\omega(j_1)
\delta(j_1+j_2+1)]\delta(m_1+m_2)
\label{2pfw}
\eeq
Now the
conformal dimension is given by $\Delta_1=
j_1(j_1+1)\rho-m_1\omega_1-k\omega_1^2/4$. The terms $A^\omega(j)$ and
$B^\omega(j)$ can
be computed as 
\begin{eqnarray}
<{V}^{\omega_1}_{j_1,m_1} \tilde{V}^{\omega_2 (-)}_{j_2,j_2}>_- =
\int \prod_{i=1}^{s_+-1} d^2 w_i
\left\langle \gamma^{j_1-m_1}_{(0)}\beta^{2j_2}_{(1)}\prod_{i=1}^{s_+-1}
\beta_{(w_i)}\right\rangle \times c.c. \times  \nonumber \\
\times \left\langle e^{\frac{2j_1}{\alpha_+}\phi(0)}e^{-\frac{(2j_2+k)}{%
\alpha_+}\phi(1)} \prod_{i=1}^{s_+-1} e^{-\frac{2}{\alpha_+}\phi(w_i,\bar
w_i)} \right\rangle ,
\end{eqnarray}
where we have taken the conjugate vertex with respect to conditions (\ref
{carga-}), but it is easy to repeat the calculation for
$\tilde V^{\omega (0)}_{j,m}$ in (\ref{fcon0})
(using the conditions (\ref{carga0})) and
the result is identical. We are omittting the $X$ and $Y$ exponentials
since
other than completing the conformal weight and determining the
conservation laws, the result does not depend on
these contributions.

For the sake of simplicity we are using the screening operator ${\cal
S_+}$.
The conservation laws ${\cal C}^{(-)}$ take now the form 
\begin{equation}
s_+=j_1-j_2 \quad , \quad m_1+j_2=0 \quad , \quad \omega_1+\omega_2=0
\end{equation}
(again we take $m_i=\bar m_i$).

Clearly it is now $A^\omega(j)$, $i.e.$ the term proportional to
$\delta(j_1-j_2)$, the one that requires no screening operators.
 In this case the contribution of the $\beta -
\gamma$ system is 
\begin{equation}
\left\langle \gamma^{j-m_1}_{(z)} \beta^{2j}_{(w)}\right\rangle = \frac{%
\Gamma(2j+1)}{(z-w)^{2j}},
\end{equation}
and the correlators of the $\phi, X, Y$ exponentials reconstruct the
conformal dimension of the two point function. So in order to normalize
(\ref{2pfw}) as before, the conjugate highest weight vertex operator has
to be
defined as 
\begin{equation}
\tilde V_{j,j}^{\omega (-)} = \frac {1}{\Gamma(2j+1)^2} \beta^{2j}
\bar\beta^{2j} e^{-\frac{(2j+k)}{\alpha_+}\phi} e^{i\sqrt {\frac{2}{k}}(j-{%
\frac{k}{2}}) X} e^{i\sqrt{\frac{2}{k}}(j-{\frac{k}{2}}\omega)Y}
\end{equation}
and thus, $A^\omega(j)=1$.

To compute the term $B^\omega(j)$, the $\beta - \gamma$ system can be
treated as described in subsection 2.2, and the result is 
\begin{equation}
\left\langle \gamma^{j_1-m_1}(0) \beta^{2j_2}(1)
\prod_{i=1}^{s_+}\beta(w_i)\right\rangle =
\frac{\Gamma(-j_1+m_1+2j_2+s_+)}{%
\Gamma(-j_1+m_1)} \prod_{i=1}^{s_+} w_i^{-1}.
\end{equation}
Taking into account the antiholomorphic contribution and the $\phi$
correlator, the Dotsenko-Fateev integrals to be computed in this case are 
\begin{equation}
\int \prod_{i=1}^{s_+-1} d^2 w_i |w_i|^{-2+8j_1/\alpha_+^2}
|1-w_i|^{-4(2j_2+k)/\alpha_+^2}
\prod_{i<j}^{s_+-1}|w_i-w_j|^{-8/\alpha_+^2}
\end{equation}

The final result obtained is 
\ber
B^\omega(j) = \frac{\Gamma(j_2+m_1)^2}{\Gamma(1+2j_2)^2\Gamma(m_1-j_1)^2}
(\pi\mu\triangle(-\rho))^{s_+}
s_+\rho^2\triangle(1-s_+)
\triangle(\rho s_+) \times
\nonumber\\
\times \delta(m_1+m_2)\delta(\omega_1+\omega_2).
\eer

It is easy to show that this expression agrees with the term $B(j)$
in equation (\ref{becker}) found in
the subsection 2.2. In order to see this,  one has to replace $j_2$ in the
steps
performed to arrive at (\ref{becker}) by $-1-j_2$
(as it is clearly needed to identify the terms $B(j)\leftrightarrow
B^\omega(j)$), and $%
m_1=-j_2$,  and the equality $\Gamma(0)/\Gamma(-s) =
(-)^s \Gamma(s+1)$ has to be used. Then we can write the result more
suggestively
as
\ber
B^\omega(j) = \triangle (1+j-m) \triangle(1+j+m)
(-\pi\mu\triangle(-\rho))^{s_+}
s_+\rho^2\triangle(1-s_+) \triangle(\rho s_+)\times
\nonumber\\
\times\delta(m_1+m_2)\delta(\omega_1+\omega_2).
\eer
where $j = j_1 = -1-j_2$, $m=m_1$.

Therefore the two point functions computed in reference
\cite{becker} for the SL(2,R)/U(1) black hole background
coincide with those obtained here for string theory in AdS$_3$ 
using a modified Coulomb gas formalism.
This agreement is not surprising since the conformal properties
completely determine the theory and even if the vertex operators have a
different representation they correspond to the same state (notice that
we are in fact comparing only the SL(2,R) part of the correlators). 
Nevertheless it should be observed that this agreement confirms the
consistency of the formalism we have developed.

The computation of the two point functions can be repeated using 
screening operators ${\cal S}_-$ and the result (\ref{b-}) for $B_-$ is
also reproduced (again replacing $j_2\leftrightarrow -1-j_2$).
Precise agreement is also found with the results obtained in the previous
Section for
the three point functions conserving winding number, for example 
$<V^{\omega_1}_{j_1,m_1} V^{\omega_2}_{j_2,m_2} \tilde
V^{\omega_3 (0)}_{j_3,j_3}>_0$ or
$<V^{\omega_1}_{j_1,m_1} V^{\omega_2}_{j_2,m_2} \tilde
V^{\omega_3 (-)}_{j_3,j_3}>_-$, give (\ref{3pf+}) or (\ref{3pf}) when the
screening operators ${\cal S}_+$ or ${\cal S}_-$ respectively are
considered. In this case one has to identify $j_3\leftrightarrow -1-j_3$.
Details of the calculations are not included because they are tedious and
similar to those already described fully in the subsection 2.2. However it
is
interesting to remark again that our prescription yields results for the
two and
three point functions matching the known exact results.

The novelty comes about when computing three point functions 
violating winding number conservation, for example
\ber
\left\langle V^{\omega_1}_{j_1,m_1}(z_1) \tilde V^{\omega_2 (-)}_{j_2,j_2}
(z_2) \tilde
V^{\omega_3 (-)}_{j_3,m_3}(z_3) \right\rangle _- =
\int \prod_{i=1}^{s_+} d^2 w_i \left |
\left\langle \gamma^{j_1-m_1}_{(z_1)} \frac{\beta^{2j_2}_{(z_2)}}
{{\Gamma(1+2j_2)}}
\frac {\beta^{2j_3}_{(z_3)}}{\Gamma(1+2j_3)}
\prod_{i=1}^{s_+}\beta_{(w_i)}\right\rangle \right |^2
\times \nonumber\\
\times \left\langle e^{\frac {2j_1}{\alpha_+}\phi(z_1,\bar z_1)}
e^{-\frac{(2j_2+k)}{\alpha_+}\phi(z_2,\bar z_2)}
e^{-\frac{(2j_3+k)}{\alpha_+}\phi(z_3,\bar z_3)}\prod_{i=1}^{s_+}
e^{-\frac{2}{\alpha_+}\phi(w_i,\bar w_i)}
\right\rangle
\eer

The exponentials of $X$ and $Y$ are not explicitly included since they
just complete
the conformal weight of the correlator, but we consider their contribution
to the conservation laws which are now given by (see (\ref{carga-}) and
(\ref{win}))
\beq
s_+ = j_1-j_2-j_3-\frac k2 \quad , \quad m_1+j_2 -\frac k 2 +j_3 = 0 \quad
, \quad \omega_1+\omega_2+\omega_3 = -1
\eeq
 
In order to compute the correlators we fix the points as usual at $z_1=0,
z_2=1$ and $z_3=\infty$.
The $\beta - \gamma$ correlator gives
\beq
\left\langle \gamma^{j_1-m_1}_{(0)} \beta^{2j_2}_{(1)}
\beta^{2j_3}_{(\infty)}\prod_{i=1}^{s_+}\beta_{(w_i)}\right\rangle 
= \frac{\Gamma(2j_2+2j_3+s_+-j_1+m_1)}{\Gamma(-j_1+m_1)}\prod_{i=1}^{s_+}
w_i^{-1} ,
\eeq
and the Dotsenko Fateev integral to be computed after performing the
$\phi$ contractions is
\beq
\int \prod_{i=1}^{s_+} d^2 w_i |w_i|^{-2-4j_1\rho}
|1-w_i|^{4\rho(j_2+\frac k2)} \prod_{i<j}|w_i-w_j|^{4\rho}
\eeq
Evaluating the integrals and simplifying the products of the
$\triangle$-functions using (\ref{fung}), the result is
\ber
{\cal A}^{j_1,j_2,j_3}_{m_1,j_2,j_3} (\delta\omega=-1) =
\frac{\Gamma(j_2+j_3-\frac k2 +
m_1)^2}{\Gamma(-j_1+m_1)^2
\Gamma(2j_2+1)^2\Gamma(2j_3+1)^2}
\times
\nonumber\\
\times (k-2)
(\pi\mu\triangle(-\rho))^{s_+} D^\omega(j_1,j_2,j_3,k)
\eer
where 
\ber
D^\omega(j_1,j_2,j_3,k) & = & D(j_1,-j_2-\frac k2,-1-j_3,k)  \nonumber\\
& = & \frac{G(j_1-j_2-j_3-\frac k2)}{G(-1)}
\frac{G(-1-j_1-j_2-\frac k2-
j_3)}{G(-2j_1-1)}\times \nonumber\\
&& \times \frac{G(1+j_1-j_2+j_3-\frac
k2)}{G(1+2j_3)}\frac{G(-j_1-j_2-\frac k2 +j_3)}{G(-2j_2-k)}
\eer

Notice that the quotient of $\Gamma -$ functions coming from the
multiplicity of the $\beta - \gamma$ correlator and the normalization of
the vertex operators can be written in a similar fashion as the
expressions obtained previouly using the equality $\Gamma(0)/\Gamma(-s)
= (-)^s \Gamma(s+1)$. Then finally we obtain
\ber
{\cal A}^{j_1,j_2,j_3}_{m_1,j_2,j_3} (\delta\omega=-1) =
\triangle (1+j_1-m_1) \triangle (-2j_2) \triangle (-2j_3)\times
\nonumber\\
\times (k-2)(-\pi\mu\triangle(-\rho))^{s_+}
 D^\omega(j_1,j_2,j_3,k).
\label{om-1}
\eer
Recall that the vertex operators with quantum numbers ($j_2,m_2$) and 
($j_3,m_3$) create conjugate highest weight states. Thus the
identification $j_2 \rightarrow
-1-j_2$ (similarly for $j_3$) is convenient in the $\triangle$-functions
above in order to compare 
this expression with (\ref{3pf+}).

Similar results are obtained if the conjugate vacuum state $\tilde
V^{\omega (0)}_{j,j}$ is
considered instead of $\tilde V_{j,j}^{\omega (-)}$. Moreover the same
expression is found if the
screening operators ${\cal S}_-$ are taken, as long as one replaces $\rho
\rightarrow \tilde\rho$ and $s_+\rightarrow s_-$.

The three point function violating winding number by one unit, equation 
(\ref{om-1}), or the equivalent one obtained replacing $\rho \rightarrow
\tilde\rho$,
presents poles given by the four $G(x)$ functions in the numerator. These
poles are located at
\ber
j_1-j_2-j_3-\frac k2 = n+m(k-2) \quad , \quad 1+j_1-j_2+j_3-\frac k2 = n +
m (k-2) \nonumber \\
-1-j_1-j_2-\frac k2 -j_3=n+m(k-2) \quad , \quad -j_1-j_2-\frac k2 +j_3 = n
+ m (k-2)
\eer
where $(n,m) \in Z^2_{\ge 0}$ or $(n,m) \in Z^2_{< 0}$. Like in the case
of winding conserving processes some of these poles are outside the
unitarity bound $-\frac 12 < j < \frac {k-3}{2}$. The remaining poles are
of the same sort as those considered pathologies in reference \cite{gk}.
 
 \section{Summary and conclusions}

In this paper we reviewed the computation of two and three point tachyon
amplitudes in the free field Coulomb gas approach to string theory in the
background of the
SL(2,R)/U(1) black hole. Two screening operators were
considered, ${\cal S}_-$ and ${\cal S}_+$. We showed that the results
obtained using
${\cal S}_+$, originally performed in reference \cite{becker}, completely
agree with previous calculations performed by
other methods. This is surprising because the Wakimoto representation 
 is expected in principle to provide a good description of the
theory when $\phi\rightarrow \infty$ ($i.e.$ far from the tip of the
cigar) where the string coupling goes to
zero, but in fact these results show that this approximation encodes the
information about 
the full theory, at least up to the three point functions. This
observation
was made before in reference \cite{satoh}
where the free field approach was carried out using a different
formalism. 
In particular, the non-perturbative term
$\Gamma(1-\frac{2j+1}{k-2})$
appears 
in this approximation. This term,  which is a finite $k$ effect, 
was observed in references \cite {zamo, kkk} 
to give rise to the same poles that appear in the Sine-Liouville
theory  which was conjectured to be the
S-dual of the SL(2,R)/U(1) coset theory.

The outcome of the Coulomb gas calculation examined in Section 2 is
unexpected also because some expressions are highly formal.
Indeed it is difficult to make sense of non-integer powers of $\beta$
fields in the correlators, so we have proceeded as if there were a
positive integer number. This is a non trivial step, eventually
justified in the light of the results accomplished, and it indicates that
the
analytic continuation in $s_+$ is well defined. 

An alternative representation of the correlators is obtained when
considering the screening operators ${\cal S}_-$. We have shown that the
results obtained in this framework match the previous ones if the
following relation holds
\beq
\pi\tilde\mu\triangle(-\tilde\rho) = (\pi\mu\triangle(-\rho))^{-\rho^{-1}}.
\eeq
An equivalent expression was recently found within the context of
Liouville
theory in reference
\cite{tesch3}, where it was interpreted as 
reflecting the self duality of the theory. In that case, additional poles
in the three point functions were found when inserting two interaction
terms in the path integral that are related by $\rho \leftrightarrow
\rho^{-1}$. Instead,
the screening operators considered here, ${\cal
S}_+$ and ${\cal S}_-$, satisfy another relation, namely ${\cal S}_- 
= ({\cal S}_+)^{-1/\rho}$, which is also verified by the Liouville
perturbations, but unlike Liouville theory the SL(2)/U(1)
coset model is not
obviously self dual. In fact, as it was mentioned above, this model was
conjectured to be 
related by a
strong/weak coupling duality symmetry to 
 the Sine-Liouville model \cite{zamo, kkk}.
Strong and weak
coupling correspond in this context to the limits $k\rightarrow\infty$ and
$k\rightarrow
2$ and viceversa. Notice that ${\cal S}_+$ and ${\cal S}_-$ respectively
can be considered as small perturbations in these regimes. 

Let us repeat that 
the conclusions to be 
drawn from these observations are not clear to us. At first sight it
seems that each screening
operator is adequate to work in a different curvature region (recall that 
$k$ is related to the radius of curvature of spacetime). However, since
the coset theory presents so many similarities with Liouville theory, self
duality might not be a priori an exception.

One hint about the resolution of these issues could be given by the four
point functions.
Some of them
were computed in the free field approximation in reference
\cite{satoh}, and the results obtained were shown to  solve the
Knizhnik-Zamolodchikov equation.
However computations are not yet available in the general case, where non
trivial
singularities are expected in the limit $z\rightarrow x$. 

We also performed the computation of the two and three point functions 
 in string theory in AdS$_3$. This
theory
was considered as the SL(2)/U(1) coset times the state space of a timelike
free boson.   
The extension of the formalism developed by Dotsenko \cite{dotse} for
SU(2) CFT proved to be adequate to explicitly introduce the spectral flow
symmetry and winding number. Conjugate vertex operators in
the correlators allowed to define scattering processes
violating winding number conservation. The results obtained for the two
and three point
tachyon amplitudes conserving winding number exhibited exact coincidence
with previous
results. This was expected due to the conformal nature of the theory.
However it is interesting to stress that this agreement also confirms the
consistency of the prescription developed in
reference \cite{gn2}.
 
Regarding this question it is interesting to notice that the conjugate
operators have
been defined for highest weight states.
It is easy to generalize the procedure applied here
to include vertex operators creating more general states in the discrete
representation. This can be done by acting on the correlators with the
currents
$J^-$ as was shown in reference \cite{becker}.
However the results obtained
apply also to states belonging to the continuous representation as it is
indicated by the equivalence of the present results with correlators
obtained by
other approaches. Actually this is a relevant case in string
theory, where one can define a notion of S-matrix for the long strings
which are precisely in the spectral flow of the continuous representation.
As it is described in reference \cite{mo} asymptotic states consisting of
long strings can
approach the center of AdS$_3$ and  scatter back to the boundary. A non
trivial result is the fact that in
this process the winding number could in principle change.
    Our prescription allows
to compute N$-$point functions violating winding number conservation for
up
to N$-2$ units. We have presented here the results for 
three point tachyon amplitudes violating winding number by one unit and
the pole structure of these expressions has been analysed.

Many open problems remain. On one hand, the computation of four point
functions is crucial for many reasons. 
These
expressions are needed to finally answer the question of the closure of
the spectrum among the unitary representations.
Moreover it
would be interesting to see if the free field approximation 
gives the correct result also for higher point functions.
 However several subtleties appear in higher point correlators, some of
which have been discussed elsewhere.

Closely related to this there is another important issue to be clarified
referring 
to the relevance of the
screening operator ${\cal S}_-$ and the physical interpretation it can
be given in the theory.

Furthermore it would be important to investigate how these issues
reflect in
the conjectured dual CFT. Additionally
the supersymmetric extension of the formalism is an interesting
problem in its own.

\bigskip
{\bf Note added in proof}: After this paper appeared in hep-th
related issues were considered in references \cite{ponjas}.

\bigskip
{\bf Acknowledgements}: 
The authors are grateful to J.
Maldacena for interesting discussions. This work is supported by grants
from CONICET (PIP98 0873) and Agencia Nacional de Promoci\'on
Cient\'{\i}fica y Tecnol\'ogica (03-03403), Argentina.

\end{document}